\documentclass[11pt,twoside]{article}
\usepackage{graphicx,epsfig,natbib,epstopdf}
\usepackage{CS18}

\begin{document}
%
%
%
\title{The MEarth-North and MEarth-South transit surveys: searching
  for habitable super-Earth exoplanets around nearby M-dwarfs}
%
%
\author{Jonathan~M.~Irwin$^{1}$, Zachory~K.~Berta-Thompson$^{2}$,
  David~Charbonneau$^{1}$, Jason~Dittmann$^{1}$,
  Emilio~E.~Falco$^{1}$, Elisabeth~R.~Newton$^{1}$, and Philip~Nutzman$^{1}$}
\affil{$^1$Harvard-Smithsonian Center for Astrophysics, 60 Garden St.,
  Cambridge, MA, USA 02138}
\affil{$^2$Massachusetts Institute of Technology, 77 Massachusetts
  Avenue, Cambridge, MA, USA 02139}
\begin{abstract}
%
%
Detection and characterization of potentially habitable Earth-size
extrasolar planets is one of the major goals of contemporary
astronomy.  By applying the transit method to very low-mass M-dwarfs,
it is possible to find these planets from the ground with present-day
instrumentation and observational techniques.  The MEarth project is
one such survey with stations in both hemispheres: MEarth-North at the
Fred Lawrence Whipple Observatory, Mount Hopkins, Arizona, and
MEarth-South at Cerro Tololo Inter-American Observatory, Chile.  We
present an update on recent results of this survey, for planet
occurrence rates, and interesting stellar astrophysics, for which our
sample of 3000 nearby mid-to-late M-dwarfs has been very fruitful.
All light curves gathered during the survey are made publicly
available after one year, and we describe how to access and use these
data.
\end{abstract}
%
%
%
%
%
\section{Introduction}

It is now widely recognized that M-dwarfs are extremely advantageous
targets to search for transiting exoplanets, due to their small sizes,
which greatly enhance transit depths, and their low luminosities,
which mean the habitable zones are at much shorter orbital periods
than for solar-type stars
(e.g. \citealt{1993Icar..101..108K,2007arXiv0706.1047C,2013ApJ...765..131K}).
For mid-to-late M-dwarfs, these factors are sufficient to allow
mini-Neptune, super-Earth and potentially even Earth-size planets to
be detected from the ground, and followed up to characterize the
planetary atmospheres spectroscopically using present day and
near-future facilities such as HST, {\it Spitzer}, JWST, and
ground-based large and extremely large telescopes.

MEarth is a dedicated ground-based transit survey designed to take
advantage of these properties, operating from two sites: the Fred
Lawrence Whipple Observatory on Mt Hopkins, Arizona, and Cerro Tololo
Inter-American Observatory, Chile.  Each site has eight $0.4$m robotic
telescopes.  MEarth-North has been fully operational since 2008
September, and MEarth-South since 2014 January.  MEarth has discovered
one transiting planet so far, the mini-Neptune GJ~1214b
\citep{2009Natur.462..891C}, which orbits an M4.5 dwarf
\citep{1995AJ....110.1838R} at $14.6$pc \citep{2013A&A...551A..48A}.
This is the smallest exoplanet for which atmospheric transmission
spectra have been obtained (e.g. \citealt{2014Natur.505...69K}) by
virtue of its small (approximately $0.2\ {\rm R}_{\odot}$) and nearby
host star.

\section{Target stars}

The MEarth survey strategy and MEarth-North target selection have been
described in detail by \citet{2008PASP..120..317N}.  All of the
targets for MEarth-North were drawn from the L{\'e}pine-Shara
Proper Motion catalog (LSPM; \citealt{2005AJ....129.1483L}),
specifically a subset of these stars with colors (and where
available, astrometric, photometric or spectroscopic distance
estimates) consistent with being M-dwarfs within $33\ {\rm pc}$ from
\citet{2005AJ....130.1680L}.

The target list for MEarth-South is still expanding at the time of
writing.  The initial set of targets were drawn from the sample of
stars with measured astrometric parallaxes from the Research
Consortium on Nearby Stars (RECONS) parallax program\footnote{{\tt
    http://www.recons.org/publishedpi.2012.1016}}, the
Palomar/Michigan State University (PMSU) spectroscopic survey
\citep{1995AJ....110.1838R,1996AJ....112.2799H}, and the LSPM-South
catalog (L{\'e}pine, private communication).  For the LSPM-South,
candidate nearby M-dwarfs were selected using the color and reduced
proper motion criteria from \citealt{2011AJ....142..138L}, except
without the $J$ magnitude limit (applying it would severely limit the
number of mid-to-late M-dwarfs selected).  The $33\ {\rm pc}$ volume
limit of the North was duplicated in the Southern target selection.

\begin{figure}
\plotone{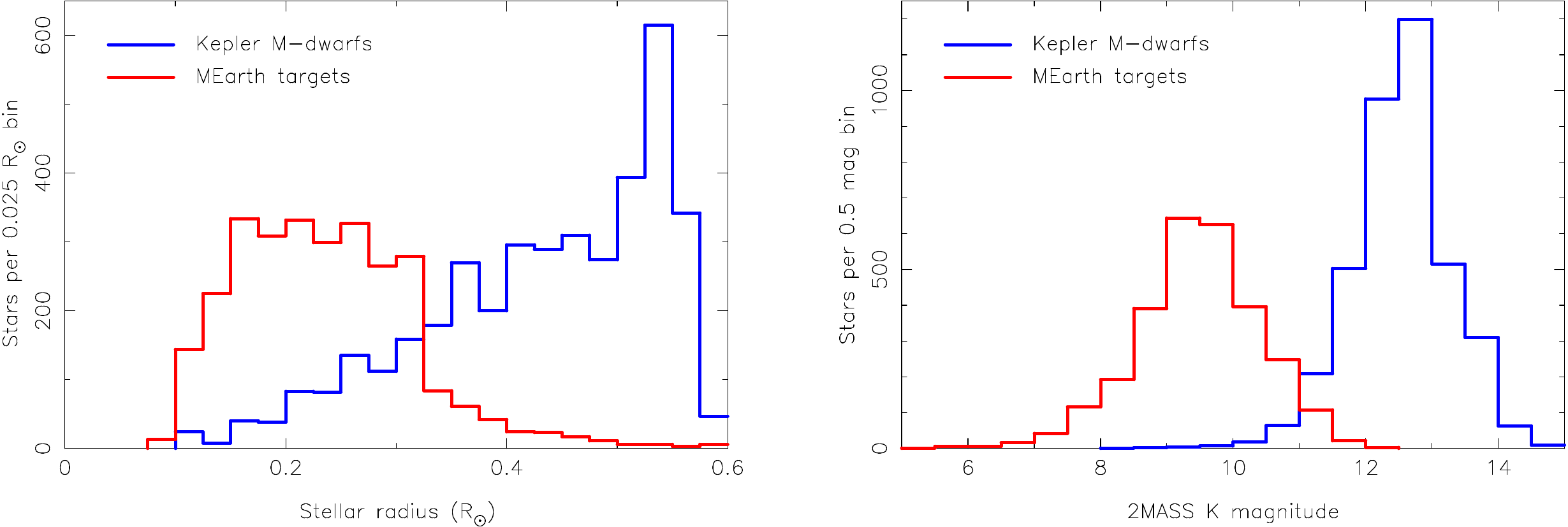}
\caption{Comparison of the properties of the MEarth target stars
with the {\it Kepler}\ M-dwarfs (parameters for the latter are from
\citealt{2013ApJ...767...95D}).  {\bf Left:} Distribution of 
estimated stellar radii.  {\bf Right:} 2MASS K magnitudes.}
\label{mearth-vs-kepler}
\end{figure}

Unlike most other surveys searching for planets around M-dwarfs,
MEarth {\it exclusively} targets mid-to-late types with estimated
radii $< 0.33\ {\rm R}_{\odot}$ (approximately M3 or later) because
these are by far the most advantageous targets to search for small
planets near the habitable zone.  The distribution of estimated
stellar radii and magnitudes for the MEarth targets is compared with
the {\it Kepler}\ M-dwarf sample in Figure \ref{mearth-vs-kepler}.  Light
curves are also available for a handful of earlier stars which lie in
the same field of view as one of our targets, or due to revisions in
the stellar parameter estimates since the original target selection.

There are approximately $2000$ targets meeting our selection criteria
in the Northern sample, and $1000$ in the current Southern sample.

\section{Observational strategy}

The MEarth targets are spread over the entire sky, meaning most of
them have to be observed individually.  During normal survey
operations, each telescope cycles around a set of target stars,
returning to each star at a cadence of $20-30$ minutes.  Exposure
times are set individually to achieve sensitivity to a desired planet
size, taking multiple exposures per visit where necessary.

\begin{figure}
\plotone{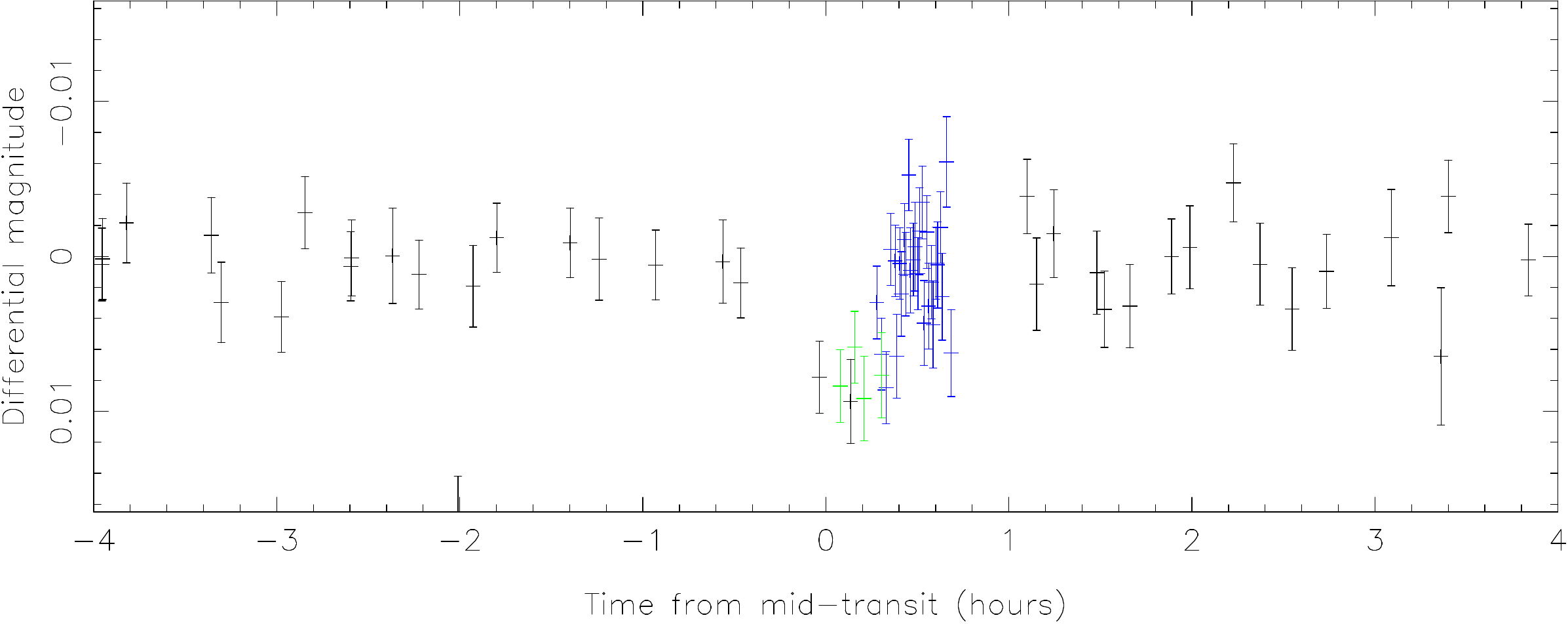}
\caption{Example phase-folded light curve of a transiting planet
  candidate discovered using our real-time detection method at
  MEarth-South. Points taken in response to the real-time detection of
  the transit are shown in green and blue, and consist of two
  separate events on different nights.  These greatly amplified the
  significance of the detection and provided an estimated orbital
  period of approximately $8$ days.}
\label{cand-rt}
\end{figure}

Transits last $0.5 - 2$ hours given the estimated parameters of our
target stars, so only a few data points would be gathered during
each transit with this strategy.  Instead, the data are analyzed in
real time, and followup observations taken immediately on any star
showing a candidate transit, while the event is still in progress
(see Figure \ref{cand-rt}).  This results in a substantial improvement
in sensitivity to long orbital periods, and maximizes use of our
telescope resources by allowing a lower cadence to be used to observe
more targets simultaneously.  Our methods for transit detection are
detailed in \citet{2012AJ....144..145B}.

\section{Planet occurrence rates}

Results from the {\it Kepler}\ mission indicate planets are extremely
common around early-M dwarfs \citep{2013ApJ...767...95D}.  Our results
for mid-to-late M-dwarfs have been shown to be consistent with the
{\it Kepler}\ results by \citet{2013ApJ...775...91B}, who also examined
methods to improve the survey yield by probing smaller planets at
longer orbital periods. These methods are undergoing trials at
MEarth-North.

\section{Stellar astrophysics}

The MEarth data and real-time transit detection system are extremely
sensitive to low-mass eclipsing binaries (EBs), out to quite long
orbital periods.  So far, $6$ new low-mass EBs have been discovered:
$4$ in the Northern hemisphere
(\citealt{2009ApJ...701.1436I,2010ApJ...718.1353I,2011ApJ...742..123I},
and one further new system), and $2$ in the Southern hemisphere
(Dittmann et al., in preparation).  These systems help to constrain
models of low-mass stars by providing valuable empirical benchmarks at
a wide range of orbital periods ($0.77-41$ days).

Rotation periods for 41 Northern stars with existing astrometric
parallaxes, based on 2 years of MEarth photometry, were used to study
rotational evolution in the fully convective domain
\citep{2011ApJ...727...56I}.  Additional data taken since the
preparation of this original sample, the expansion to the Southern
hemisphere, and recent improvements in the availability of astrometric
parallax measurements allow this type of analysis to be extended to a
much larger sample of stars, which is in progress (Newton et al., in
prep.).

A spectroscopic followup program has targeted many of the objects with
measured rotation periods, both in the optical to measure activity
indicators (West et al., in prep.) and the near-infrared to
measure metallicities \citep{2014AJ....147...20N}.  These
observations have also been used to develop an empirical method based
on equivalent widths of atomic lines in the near-infrared to
estimate fundamental stellar properties such as radii, calibrated
using observations of stars with interferometric angular diameter
measurements (Newton et al., in prep.).

Astrometric parallaxes were measured from MEarth data for $1507$
Northern hemisphere mid-to-late M-dwarfs \citep{2014ApJ...784..156D},
using the transit survey data where available, and a dedicated
observing program covering all other target stars at a cadence of
approximately $10$ days for $3$ years.  Many of these stars did not
have previous astrometric parallax measurements.

\section{Data releases}

Public releases of the MEarth target star light curves are made
annually on September 1, with data being released one year after being
taken.  At the time of writing, Data Release 3 is in the late stages of
preparation, and combined with the existing Data Release 2 (released
2013 September 1) will include all light curves from the start of the
Northern survey to 2013 July (the end of our 2012-2013 observing
season), a total of $5$ years of data.  All release materials are
placed on a public web page\footnote{{\tt
    http://www.cfa.harvard.edu/MEarth/Data.html}}.

These releases are intended to be accessible and as straightforward as
possible to use.  Light curves and summary tables of target star
properties are provided in ASCII format, including tar files
containing the entire release for those desiring to analyze large
numbers of objects.  Details of the content and production of the
light curves (including how the data were processed) are included.

\acknowledgments{
It is a pleasure to acknowledge the assistance of the staff at the
Fred Lawrence Whipple Observatory and Cerro Tololo Inter-American
Observatory in making MEarth a reality.  The MEarth Team gratefully
acknowledges funding from the David and Lucile Packard Fellowship for
Science and Engineering (awarded to D.C.).  This material is based
upon work supported by the National Science Foundation under grants
AST-0807690, AST-1109468, and AST-1004488 (Alan T. Waterman Award).
This publication was made possible through the support of a grant from
the John Templeton Foundation.  The opinions expressed in this
publication are those of the authors and do not necessarily reflect
the views of the John Templeton Foundation.
}

\end{document}